\documentclass[12pt,revtex4]{article}
\usepackage{graphicx}

\begin{document}
\textwidth=135mm
 \textheight=210mm
\begin{center}
{\bfseries Update on the status of Hadronic Squeezed Correlations at RHIC Energies
\footnote{{\small Talk presented at the VI Workshop on Particle Correlations and Femtoscopy (WPCF 2010), Bogolyubov Institute, Kiev, Ukraine, September 14 - 18,
2010.}}}
\vskip 5mm
Sandra S. Padula$^{\ddag}$, Danuce M. Dudek$^\ddag$, and Ot\'avio Socolowski, Jr.$^\dag$ 
\vskip 5mm
{\small {\sl $^\ddag$ Inst. F\'\i sica Te\'orica - UNESP, C. P. 70532-2, 01156-970 S\~ao Paulo, SP, Brazil}} \\
{\small {\sl $^\dag$ IMEF - FURG - C. P. 474, 96201-900, Rio Grande, RS, Brazil}}\\
\end{center}
\vskip 5mm
\centerline{\bf Abstract}
In high energy heavy ion collisions a hot and dense medium is formed, where the 
hadronic masses may be shifted from their asymptotic values. If this mass modification occurs, 
squeezed back-to-back correlations (BBC) of particle-antiparticle pairs are 
predicted to appear, both in the femionic (fBBC) and in the bosonic (bBBC) sectors. Although they have unlimited intensity even for finite-size expanding systems, these hadronic squeezed correlations are very sensitive to 
their time emission distribution. Here we discuss results in case this time emission is parameterized  
by a L\'evy-type distribution, showing that it reduces the signal even more dramatically than a Lorentzian distribution, which already reduces the intensity of the effect by orders of magnitude, as compared to the sudden emission. However, we show that the signal could still survive if the duration of the process is short, and if the effect is searched for lighter mesons, such as kaons. We compare some of our results to recent PHENIX preliminary data on squeezed correlations of $K^+K^- $ pairs. 

\vskip 10mm

The history of the squeezed correlations between particle-antiparticle pairs started in the early 1990's\cite{weiner} but its final formulation was proposed almost a decade later, by M. Asakawa et al. \cite{acg99}. In such approach, the  {\sl squeezed back-to-back correlations} (BBC) of boson-antiboson pairs resulted from a quantum mechanical  unitary transformation relating in-medium quasi-particles to two-mode squeezed states of their free counterparts, originated in a modification of the particles' mass in the hot and dense medium. Shortly after that, P. K. Panda et al.\cite{pkchp01} showed that a similar BBC between fermion-antifermion pairs should exist,  if the masses of these particles were modified in-medium. Both the fermionic (fBBC) and the bosonic (bBBC)  back-to-back squeezed correlations are described by analogous formalisms, being both positive correlations of unlimited intensity. In the remainder of this paper, we focus our discussion into the case of charged kaons, i.e., $K^+$ and $K^-$ pairs, for which the squeezed correlations are predicted. Since in this case there is no contribution from identical particles, 
the correlation function 
is written as
$C_s({\mathbf k}_1,{\mathbf k}_2)  =
1+ \frac{| G_s(1,2) |^2}{G_c(1,1) G_c(2,2) }$, 
where 
$G_c(i,i) =\omega_{\mathbf k_i}\,
\langle \hat{a}^\dagger_{\mathbf k_i} \hat{a}_{\mathbf k_i} \rangle = \!\omega_{\mathbf k_i} \frac{d^3N}{d\mathbf k_i}$, is the spectrum of each particle ($i=1,2$) and 
$G_s(1,2) =
\sqrt{\omega_{{\mathbf k}_1} \omega_{{\mathbf k}_2} }\langle \hat{a}_{
{\mathbf k}_1} \hat{a}_{{\mathbf k}_2} \rangle $ is the squeezed amplitude, which gives a non-zero contribution if 
the hadronic masses are modified in-medium. 
This non-vanishing contribution to $\langle \hat{a}^{(\dagger)}_{
{\mathbf k}_1} \hat{a}^{(\dagger)}_{{\mathbf k}_2} \rangle $ is originated in the Bogolyubov-Valatin transformation relating the asymptotic operators, $\hat{a}_k$  ($\hat{a}^\dagger_k$),  to their in-medium counterparts, $\hat{b}_k$ ($\hat{b}^\dagger_k$), i.e., $a_k=c_k b_k + s^*_{-k} b^\dagger_{-k} \; ; \; a^\dagger_k=c^*_k
b^\dagger_k + s_{-k} b_{-k}$, where $c_k=\cosh(f_k)$, $s_k=\sinh(f_k)$. The argument,  
$f_{i,j}(x)=\frac{1}{2}\log\left[\frac{K^{\mu}_{i,j}(x)\, u_\mu
(x)} {K^{*\nu}_{i,j}(x) \, u_\nu(x)}\right]$, is the  {\sl squeezing parameter}, where $K^{\mu}_{i,j}(x)=\frac{1}{2} (k_i^\mu+k_j^\mu) $ is the average of the momenta of each particle, and $u_\mu$ is the flow velocity of the system. The details of the formalism and relations used to produce the results shown here were derived in Ref. \cite{acg99,pkchp06,dansan10,sanota10}. 


In Ref.\cite{acg99,pkchp01}, the case of static infinite medium was considered and later, in Ref.\cite{pkchp06,dansan10,sanota10}, a finite system expanding with moderate radial flow was analyzed. For simplicity, 
the expansion was considered to be non-relativistic and the squeezing parameter, to be flow and momentum-independent.  A constant mass shift was assumed, homogeneously distributed over the system or in part of it, and linearly related to the asymptotic mass, i.e., $m_* =m \pm \delta M$ \cite{pkchp01}-\cite{sanota10}.These simplifications allowed to obtain analytical expressions for the BBC correlation functions, whose forms were detailed in Ref.\cite{pkchp06,dansan10,sanota10}. Several interesting features of the squeezed correlations were discussed in those references. 
One of them, in particular, is rather striking, i.e., the time emission distributions play a crucial r\^ole in the possible observation of the squeezing correlations. In previous works, it was shown that the strength of the squeezed correlation could be very high if the emission was instantaneous, i.e., described by a $\delta$ function at freeze-out. In the case of $K^+ K^-$ pairs this could reach values as high as $C(k,-k) \sim 330$, at the maximal mass shift, $m_* = 350$ MeV, for expanding systems with radial flow velocity, $<u>=0.5$, as illustrated in Fig. 1(a).  
\begin{figure}[h]
\begin{center}
 \includegraphics[width=.5\textwidth]{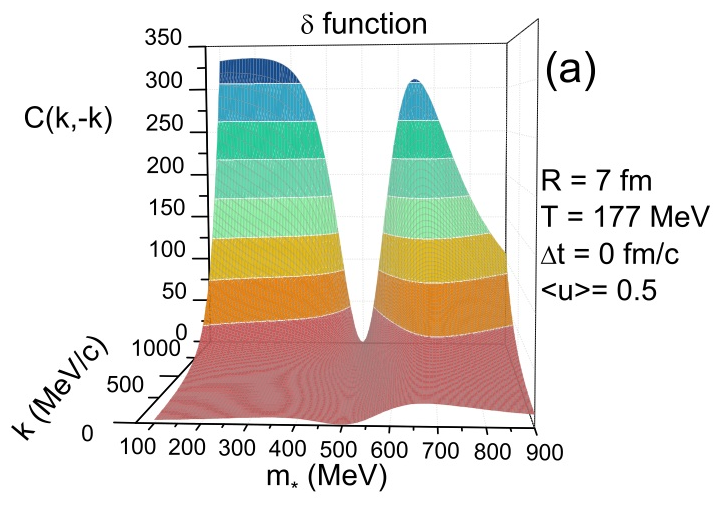}\includegraphics[width=.5\textwidth]{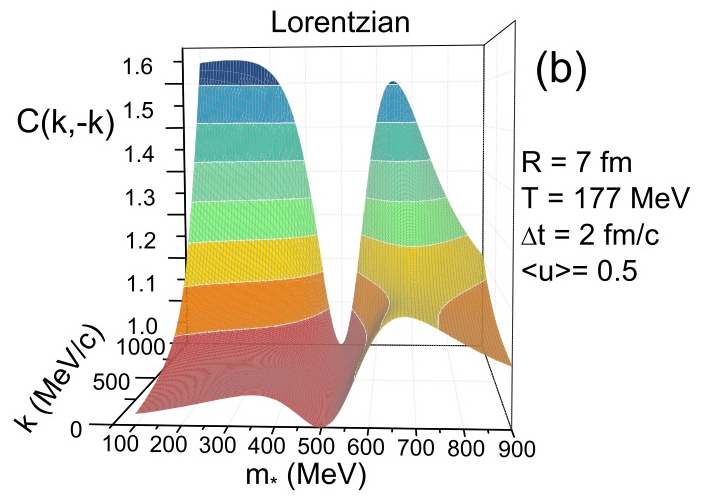} \includegraphics[width=.5\textwidth]{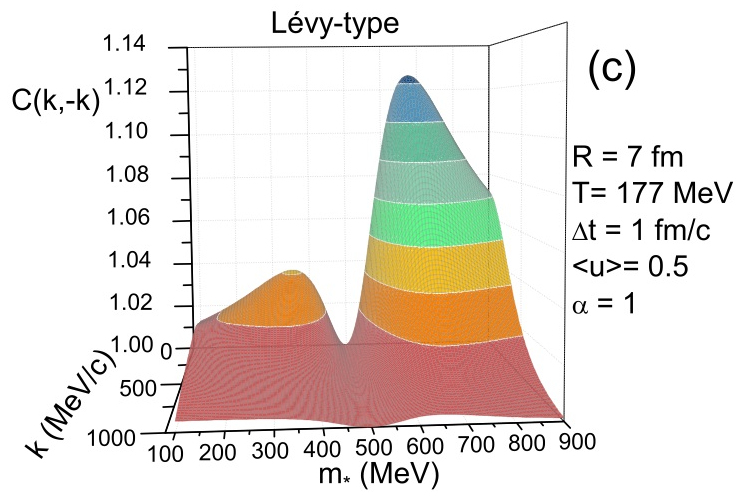} \includegraphics[width=.5\textwidth]{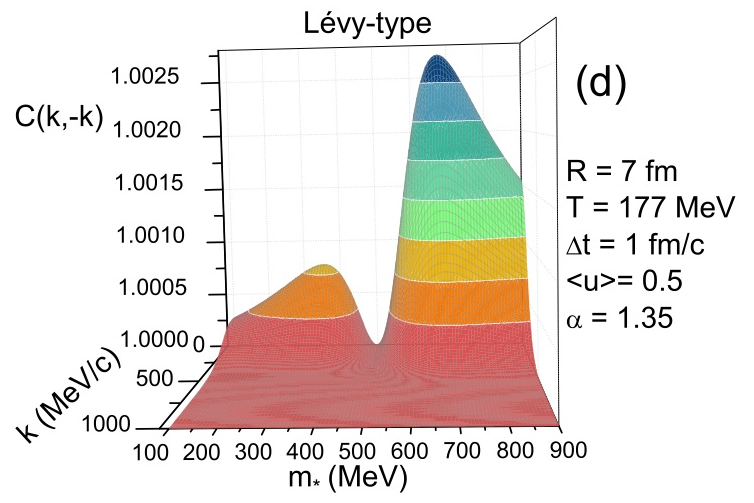}
\caption{The squeezed to correlation function in terms of the modified mass, $m_*$,  and the relative momentum of the pair, $k=|\vec{k}|$, is shown for different time distributions. Part (a) shows the results for the case of a $\delta$ function in time;  part (b) shows results for a Lorentzian with $\Delta t=2$ fm/c; (c), for a L\'evy distribution with $\alpha=1$ and (d), for  $\alpha=1.35$, both for $\Delta t=1$ fm/c.}\label{fig1}
\end{center}
\end{figure}
 However, if the emission lasts a finite interval, the squeezed term should be multiplied by a factor which may reduce this intensity considerably. For instance, for a Lorentzian time emission, $|{\cal F}(\Delta t)|^2=[1+(\omega_1+\omega_2)^2 \Delta t ^2]^{-1}$, with a time interval $\Delta t =2$ fm/c, as adopted in Ref.\cite{acg99}-\cite{sanota10}, the maximum intensity of the squeezed correlation function may be 
 roughly a factor of 20 smaller than in the case of instantaneous emission. This is shown in Fig. 1(b). 
Nevertheless, since we do not know a priori which functional form would be privileged by nature, we also investigate the time emission as a symmetric, $\alpha$-stable L\'evy distribution, 
$|F(\Delta t)|^2=\exp\{-[\Delta t (\omega_1+\omega_2)]^\alpha\}$. 
This functional form was used to fit PHENIX data\cite{levy-phenix} on two- and three-particle Bose-Einstein correlations. The distribution index was found to be $\alpha \sim 1$, for $0.2 < m_T < 0.3$ GeV, and $\alpha \sim 1.35$, for $0.3 < p_T < 0.45$ GeV/c, for which we investigate the time emission factor. 
This is shown in Fig. 1(c) and (d), respectively. We see that the L\'evy distribution reduces the intensity of the squeezed correlation function even more dramatically than the Lorentzian distribution. 

The results shown in Fig.\ref{fig1} are important but not practical for the purpose of searching for hadronic squeezed correlations empirically,  since the modified mass, $m_*$, is not observable  outside  the hot and dense medium, and the particle and antiparticle momenta are hardly measured as exactly back-to-back. Instead, we should consider distinct values for the momenta of the particles, ${\mathbf k_1}$ and ${\mathbf k_2}$, and combine them in measurable quantities, e.g., as their average, 
${\mathbf K}_{12}\!=\!\frac{1}{2}( {\mathbf k_1}+{\mathbf k_2})$, and their relative momenta, ${\mathbf q_{12}}\!=\!( {\mathbf k_1}-{\mathbf k_2})$ \cite{pkchp06,dansan10,sanota10}. This approach considers non-relativistic momenta and therefore has its application constrained to this limit. 
For a relativistic treatment, see discussion in Ref. 
\cite{qm08}.  Naturally, the values of the squeezed correlation function in Fig. \ref{fig1} show only the maximum expected value, i.e., its  intercept at $K_{12}=0$, for different values of $|\vec{q}_{12}|=|\vec{k}_1-\vec{k}_2|=|\vec{k}-(-\vec{k})|=2|\vec{k}|$.

\begin{figure}[h]
\begin{center}
\includegraphics[width=.5\textwidth]{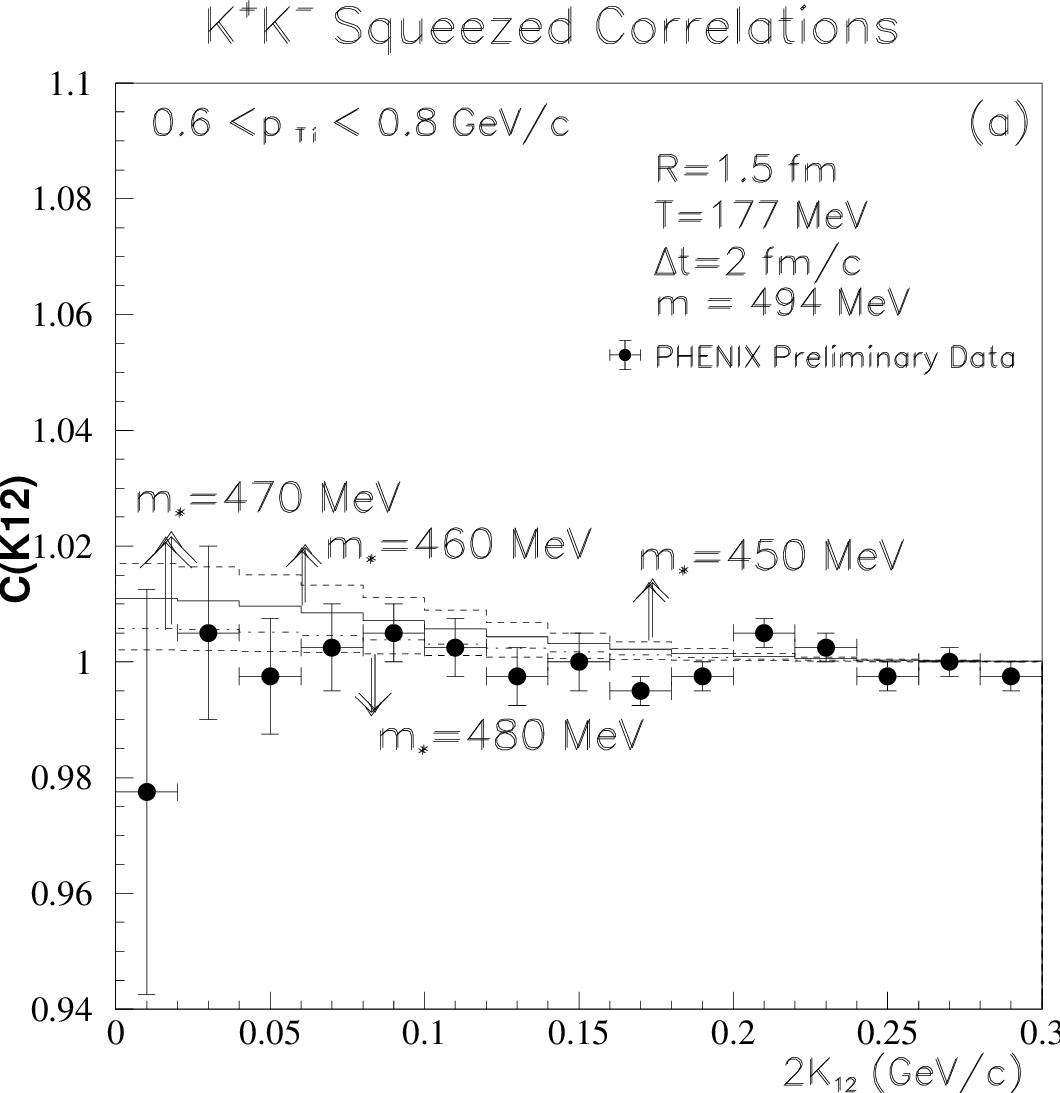}\includegraphics[width=.5\textwidth]{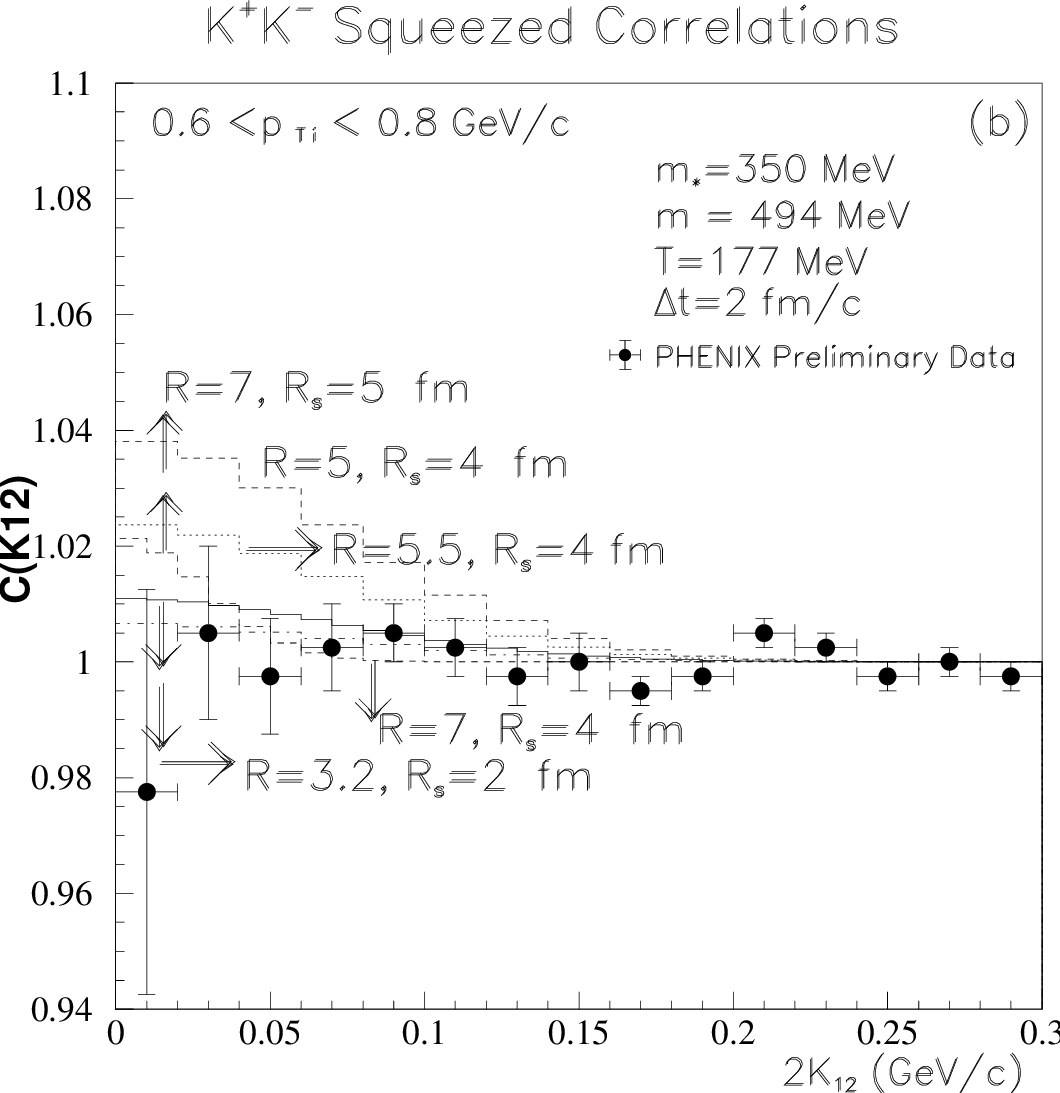} 
\caption{Part (a) shows the comparison with data of the squeezed correlation function versus $2K_{12}$, for several values of $m_*$ and $R=1.5$ fm. Part (b) shows similar results considering that the squeezing occurs in a smaller portion (with radius $R_s$) of the system of radius $R$, for different values of radii.}\label{fig2}
\end{center}
\end{figure}

In our non-relativistic model we consider that the system formed in high energy collisions is described by a Gaussian source with circular cross-sectional area of radius $R$ (see Ref.\cite{pkchp06} for details). In the studies shown in Fig. 1, the mass modification was supposed to be homogeneously distributed over all the system of radius $R=7$ fm. However, the signal of the resulting correlation function falls too fast to unity if compared to the PHENIX preliminary data on squeezed hadronic correlations, presented in Ref.\cite{mn-wpcf09}. 
In contrast, we consider in Fig. 2(a) the result of the simulation with our model, for a small system with $R=1.5$ fm, where the squeezed correlation is shown as a function of $2 K_{12}$, for particles with transverse momenta in the range $0.6 < m_T < 0.8$, as in the experiment. Results for several values of the modified mass, $m_*$, are shown. Although describing the trend of data reasonably well,  specially for $m_*=460$ MeV (continuous line), such radius is too small as compared with measurements from the PHENIX Collaboration in similar conditions (see discussion in Ref.\cite{dansan10}). 
\begin{figure}[h]
\begin{center}
\includegraphics[width=.8\textwidth]{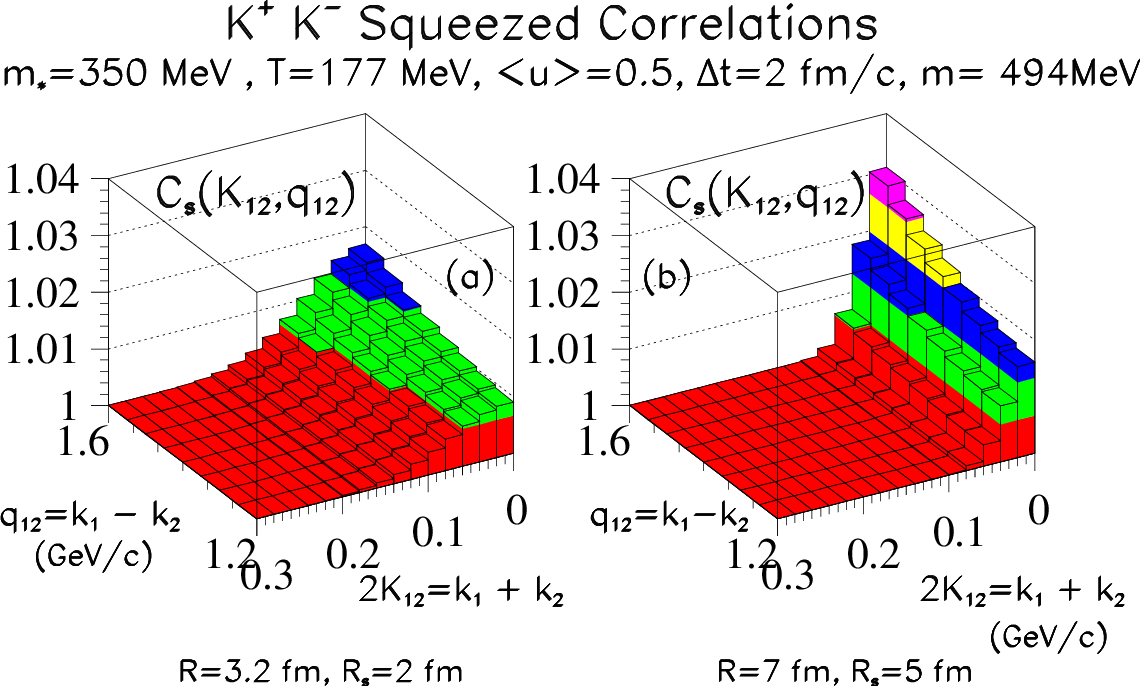}
\caption{Part (a) and (b) show $C(K_{12},q_{12}$ versus $K_{12}$ and $q_{12}$, for two sets of  parameters for two corresponding histograms in Fig. 2(b). }\label{fig3}
\end{center}
\end{figure}
Besides, also the hypothesis of mass modification over all the system may be too strong. So, we considere also  the two-volume case discussed in Ref.\cite{pkchp06}, in which the squeezing is assumed to be restricted to an smaller portion, with radius $R_s$, inside of the system of radius $R$. This is shown in Fig. 2(b)  for several combinations of the these values of radii, fixing the shifted kaon mass as $m_*=350$MeV. We see that some of the parameters in Fig. 2(b) describe the data qualitatively well, in particular the histograms corresponding to $R=7.0$ fm, $R_s=5$ fm and $R=3.2$  fm, $R_s=2$ fm, i.e., there is ambiguity in the scenarios.  Nevertheless, in case the statistics could be increased considerably, a study of the correlation function, $C(K_{12},q_{12})$, in terms of these variables in the plane $(2 K_{12},q_{12})$, could improve the distinction of  different scenarios, as shown in Fig. 3(a) and (b), for the most prominent sets of variables of Fig. 2(b). 

As a final remark, we could say that, a comparison of our two-volume non-relativistic model with the preliminary PHENIX data on the squeezed hadronic correlations of Ref.\cite{mn-wpcf09}, does not allow for unambiguously conclude in favor of in-medium mass modification at RHIC energies. Nevertheless, it does not discard it neither! Higher statistics data are necessary to help turn this crucial point into a positive or negative answer, besides solving the ambiguity of the parameters describing the system. The analysis of higher statistics data on $C(K_{12},q_{12})$ in the plane $(K_{12},q_{12})$ could improve the distinction of  different scenarios and, hopefully, allow for concluding in favor of in-medium mass modification already at RHIC energies.  

\bigskip
SSP gratefully acknowledges the support by CAPES for participating in the WPCF 2010. DMD also thanks CAPES for the support during the completion of this work. OSJ acknowledges funding from CNPQ.



\end{document}